\documentclass{sig-alternate}
\usepackage{times}
\usepackage[latin1]{inputenc}
\usepackage{amsmath}
\usepackage{amsfonts}
\usepackage{amssymb}
\usepackage{verbatim}
\usepackage{algorithm}
\usepackage{algpseudocode}
\usepackage{graphicx}
\usepackage{subfigure}
\usepackage{enumitem}
\usepackage{url}
\usepackage{balance}
\usepackage{epsfig}
\usepackage{cite}
\algnotext{EndFor}
\algnotext{EndIf}
\algnotext{EndWhile}
\begin{document}
\title{Query Optimization for Dynamic Graphs}
\numberofauthors{6}
\author{
%
%
\alignauthor
Sutanay Choudhury\\
\affaddr{Pacific Northwest National Laboratory, USA}\\
\email{sutanay.choudhury@pnnl.gov}
\alignauthor
Lawrence Holder\\
\affaddr{Washington State University, USA}\\
\email{holder@wsu.edu}
\alignauthor
George Chin\\
\affaddr{Pacific Northwest National Laboratory, USA}\\
\email{george.chin@pnnl.gov}
\and  
\alignauthor
Patrick Mackey\\
\affaddr{Pacific Northwest National Laboratory, USA}\\
\email{ patrick.mackey@pnnl.gov}
\alignauthor
Khushbu Agarwal\\
\affaddr{Pacific Northwest National Laboratory, USA}\\
\email{khushbu.agarwal@pnnl.gov}
\alignauthor
John Feo\\
\affaddr{Pacific Northwest National Laboratory, USA}\\
\email{john.feo@pnnl.gov}
}
\maketitle
\thispagestyle{empty}
\begin{abstract}
Given a query graph that represents a pattern of interest, the emerging pattern detection problem can be viewed as a continuous query problem on a dynamic graph.  We present an incremental algorithm for continuous query processing on dynamic graphs.  The algorithm is based on the concept of query decomposition;  we decompose a query graph into smaller subgraphs and assemble the result of sub-queries to find complete matches with the specified query.  The novelty of our work lies in using the subgraph distributional statistics collected from the dynamic graph to generate the decomposition.  We introduce a ``Lazy Search" algorithm where the search strategy is decided on a vertex-to-vertex basis depending on the likelihood of a match in the vertex neighborhood.  We also propose a metric named ``Relative Selectivity" that is used to select between different query decomposition strategies.   Our experiments performed on real online news, network traffic stream and a synthetic social network benchmark demonstrate 10-100x speedups over competing approaches.
\end{abstract}
\section{Introduction}





Social media streams and cyber data sources such as computer network traffic are prominent examples of high throughput, dynamic graphs.  Application domains such as computational journalism, emergency response, national security put a premium on discovering critical events as soon as they emerge in the data.  Thus, processing streaming updates to a dynamic graph database for real-time situational awareness is an important research problem.  Apart from their dynamic nature, these particular data sources are also distinguished by their heterogeneous or multi-relational nature.  For example, a social media data stream contains a diverse set of entity types such as person, movie, images etc. and relations such as (\textsl{friendship, like etc.}).  For cyber-security, a network traffic dataset can be modeled as a graph where vertices represent IP addresses and edges are typed by classes of network traffic \cite{Joslyn:2013:MSC:2484425.2484428}.  Our work is focused on continuous querying of these dynamic, multi-relational graphs.  We want to register a pattern as a graph query and continuously perform the query on the data graph as it evolves over time.



Continuous querying of a dynamic graph raises a number of unique challenges.  Indexing techniques that preprocess a graph and speed up queries are expensive to periodically recompute in a dynamic  setting.  Periodic execution of the query is an obvious solution under this condition, but the effectiveness of this approach will reduce as the interval between query executions shrinks.  Also, periodic searching of the entire graph can be wasteful where the query match emerges slowly \cite{Choudhury:2013:DyNetMM, Choudhury:2013:SSD:2463676.2463697} because we will find a partial match for the query every time we search and potentially redo the work numerous times.


The following describes the key idea behind our solution.  We approach the problem from an incremental processing perspective where search happens locally on every edge arrival.  We do not search for the entire query graph around every new edge.  Given a query graph, we decompose it into smaller subgraphs as ordered by their selectivity.  The selectivity information is obtained using the single-edge level and 2-edge path distribution obtained from the graph stream (section \ref{sec:Query Optimization}).  We store the resulting decomposition into a data structure named SJ-Tree (Subgraph Join Tree) (section \ref{sec:Incremental Query Processing}) that tracks matching subgraphs in the data graph.  For a new edge in the graph, we always search for the \textsl{most selective} subgraph of the query graph.  For other subgraphs of the query graph, a search is triggered if and only if a match for the previous subgraph in the selectivity order was obtained in the neighborhood of the new edge.  This algorithm named ``Lazy Search" is described in section \ref{sec:Lazy Search}.  We introduce two metrics, \textsl{Expected and Relative Selectivity}, that captures the effectiveness of a given query decomposition (section \ref{sec:Query Optimization}).  Further, we demonstrate how these metrics can be used to reason about the performance from different decompositions and select the best performing strategy.


\subsection{Contributions}

The most important takeaway from our work is that even as the subgraph isomorphism problem is NP-complete, it is possible to perform efficient continuous queries on dynamic graphs by exploiting the heterogeneity in the data and query graph.  More specific contributions from the paper are listed below.


\begin{enumerate}
\item We present a dynamic graph search algorithm that demonstrates speedup of multiple orders of magnitude with respect to the state of the art.
\item We introduce two selectivity metrics for query graphs that are estimated using efficiently obtainable distributional statistics of single edge and 2-edge subgraphs from the graph stream.
\item We present an automatic query decomposition algorithm that selects the best performing strategy using the aforementioned graph stream statistics and \textsl{Relative Selectivity}.
\end{enumerate}

Our observations are supported by experiments on datasets from three diverse domains (online news, computer network traffic and a social media stream).
\section{Background and Related Work}
\label{sec:background}

This section is aimed at providing an overview of the related field and provide the context for the studied problem.  We begin with introducing the key concepts.

\textbf{Multi-Relational Graphs} We define a graph $G$ as an ordered-pair $G = (V, E)$ where $V$ is the set of vertices and the $E$ is the set of edges that connect the vertices.  An edge represents a pair of vertices, also  known as \textit{end points}.  In the following, we use $V(G)$ and $E(G)$ to indicate the set of vertices and edges associated with a graph $G$.  A \textsl{labeled graph} is a six-tuple $G = (V, E, \Sigma_V, \Sigma_E, \lambda_V, \lambda_E)$, where $\Sigma_V$ and $\Sigma_E$ are sets of distinct labels for vertices and edges.  $\lambda_V$ and $\lambda_E$ are vertex and edge labeling functions, i.e. $\lambda_V : V \rightarrow \Sigma_V$ and $\lambda_E : E \rightarrow \Sigma_E$.

\textbf{Dynamic Graphs} We define \textit{dynamic graphs} as graphs that are changing over time through edge insertion or deletion.  Every edge in a dynamic graph has a timestamp associated with it and therefore, for any subgraph $g$ of a dynamic graph we can define a time interval $\tau(g)$ which is equal to the interval between the earliest and latest edge belonging to $g$.  We focus on directed, labeled dynamic graphs with multi-edges in this work.  The graph is maintained as a window in time.  Given a time window $t_W$, edges are deleted as they become older than $t_{last} - t_W$, where $t_{last}$ is the timestamp of the newest edge in the graph.  

\textbf{Continuous Queries} A continuous query can be described as computing a function $f$ over a stream $S$ continuously over time and notifying the user whenever the output of $f$ satisfies a user-defined constraint \cite{Law:2011:RLD:1966385.1966386}.  They are distinguished from ad-hoc query processing by their high selectivity (looking for unique events) and need to detect newer updates of interest as opposed to retrieving lots of past information.  In this paradigm the primary objective is to notify a listener as soon as the query is matched.  One may view conventional databases as passive repositories with large collections of data that work in a request-response model whereas continuous queries are data-driven or trigger oriented.  These features challenge many of the fundamental assumptions for conventional databases and establish continuous query processing on relational data streams as a major research area.   The literature on database research from the past two decades is abundant with work on continuous query systems \cite{Chandrasekaran:2003:TCD:872757.872857, DBLP:conf/sigmod/ArasuBBDIRW03}.  Babcock et al. \cite{Babcock:2002:MID:543613.543615} provide an excellent overview of continuous query systems and their design challenges.

\textbf{Subgraph Isomorphism} Given the query graph $G_q$ and a matching subgraph of the data graph ($G_d$) denoted as $G^{'}_d$, a matching between $G_q$ and $G^{'}_d$ involves finding a bijective function $f : V(G_q) \rightarrow V(G^{'}_d)$ such that for any two vertices $u_1, u_2 \in V(G_q)$,  $(u_1, u_2) \in E(G_q) \Rightarrow (f(u_1), f(u_2)) \in E(G^{'}_d)$.

\subsection{Problem Statement}
\label{subsec:Problem Statement}

Every edge in a dynamic graph has a timestamp associated with it and therefore, for any subgraph $g$ of a dynamic graph we can define a time duration $\tau(g)$ which is equal to the duration between the earliest and latest edge belonging to $g$.  Given a dynamic multi-relational graph $G_d$, a query graph $G_q$ and a time window $t_W$, we report whenever a subgraph $g_d$ that is isomorphic to $G_q$ appears in $G_d$ such that $\tau(g_d) < t_W$.  The isomorphic subgraphs are also referred to as \textit{matches} in the subsequent discussions.  Assume that $G^k_d$ is the data graph at time step $k$.  If $M(G^k_d)$ is the cumulative set of all matches discovered until time step $k$ and $E_{k+1}$ is the set of edges that arrive at time step $k+1$, we present an algorithm to compute a function $f\left(G_d, G_q, E_{k+1}\right)$ which returns the incremental set of matches that result from updating $G_d$ with $E_{k+1}$ and is equal to $M(G^{k+1}_d) - M(G^k_d)$.

\subsection{Related Work}
\label{sec:Related Work}
Graph querying techniques have been studied extensively in the field of pattern recognition over nearly four decades \cite{conte2004thirty}.  Two popular subgraph isomorphism algorithms were developed by Ullman \cite{Ullmann:1976:ASI} and Cordella et al. \cite{cordella2004sub}.  The VF2 algorithm \cite{cordella2004sub} employs a filtering and verification strategy and outperforms the original algorithm by Ullman.  Over the past decade, the database community has focused strongly on developing indexing and query optimization techniques to speed up the searching process.  A common theme of such approaches is to index vertices based on k-hop neighborhood signatures derived from labels and other properties such as degrees and centrality  \cite{tian2008tale, Tong:2007:FBP:1281192.1281271, Zhao:2010:GQO:1920841.1920887}.   Other major areas of work involve exploration of subgraph equivalence classes \cite{Han:2013:TIT:2463676.2465300} and search techniques for alternative representations such as similarity search in a multi-dimensional vector space \cite{Khan:2011:NBF:1989323.1989418}.  Apart from neighborhood based signatures, \textit{graph sketches} is an important area that focuses on generating different synopses of a graph data set \cite{Zhao:2011:GQE:2078331.2078335}.  Development of efficient graph sketching algorithms and their applications into query estimation is expected to gain prominence in the near future.

Investigation of subgraph isomorphism for dynamic graphs did not receive much attention until recently.  It introduces new algorithmic challenges because we can not afford to index a dynamic graph frequently enough for applications with real-time constraints.  In fact this is a problem with searches on large static graphs as well \cite{DBLP:journals/pvldb/SunWWSL12}.  There are two alternatives in that direction.  We can search for a pattern repeatedly or we can adopt an incremental approach.  The work by Fan et al. \cite{Fan:2011:IGP:1989323.1989420} presents incremental algorithms for graph pattern matching.  However, their solution to subgraph isomorphism is based on the repeated search strategy.  Chen et al. \cite{Chen:2010:CSP:1850481.1850517} proposed a feature structure called the \textsl{node-neighbor tree} to search multiple graph streams using a vector space approach.  They relax the exact match requirement and require significant pre-processing on the graph stream.  Our work is distinguished by its focus on temporal queries and handling of partial matches as they are tracked over time using a novel data structure.  From a data-organization perspective,  the SJ-Tree approach has similarities with the Closure-Tree \cite{He:2006:CIS:1129754.1129898}.  However, the closure-tree approach assumes a database of independent graphs and the underlying data is not dynamic.  There are strong parallels between our algorithm and the very recent work by Sun et al. \cite{DBLP:journals/pvldb/SunWWSL12}, where they implement a query-decomposition based algorithm for searching a large static graph in a distributed environment.  Here our work is distinguished by the focus on continuous queries that involves maintenance of partial matches as driven by the query decomposition structure, and optimizations for real-time query processing.  Mondal and Deshpande \cite{Mondal:2014} propose solutions to supporting continuous ego-centric queries in a dynamic graph,  Our work focuses on subgraph isomorphism, while \cite{Mondal:2014} is primarily focused on aggregate queries.  We view this as complementary to our work, and it affirms our belief that continuous queries on graphs is an important problem area, and new algorithms and data structures are required for its development.
\section{A Query Decomposition Approach}

\label{sec:Technical Approach}

%
%
%
%
%

We introduce an approach that guides the search process to look for specific subgraphs of the query graph and follow \textsl{specific} transitions from small to larger matches.  Following are the main intuitions that drive this approach.
\begin{enumerate}
\item Instead of looking for a match with the entire graph or just any edge of the query graph, partition the query graph into smaller subgraphs and search for them.
\item Track the matches with individual subgraphs and combine them to produce progressively larger matches.
\item Define a \textit{join order} in which the individual matching subgraphs will be combined.  Do not look for every possible way to combine the matching subgraphs.
\end{enumerate}

\begin{figure}[]
\includegraphics[scale=0.4]{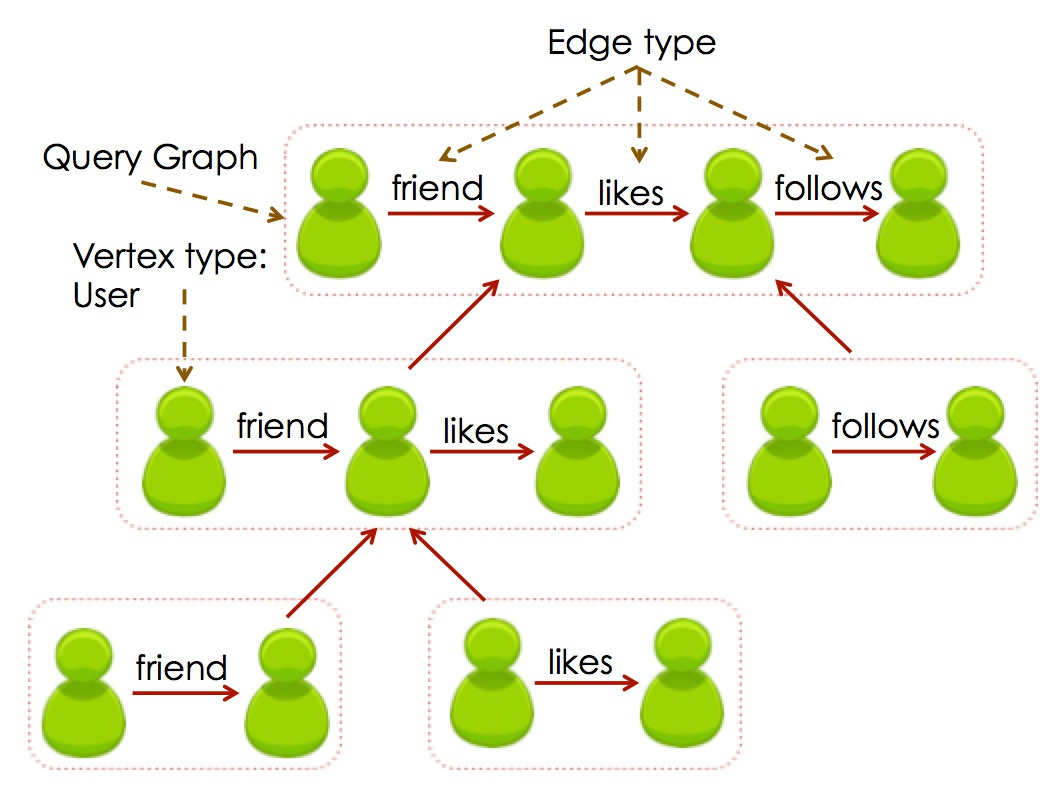}
\caption{Illustration of the decomposition of a social query in SJ-Tree. }
\label{fig:join_tree}
\end{figure}

Figure \ref{fig:join_tree} shows an illustration of the idea.  Although the current work is completely focused on temporal queries, the graph decomposition approach is suited for a broader class of applications and queries.  The key aspect here is to search for substructures without incurring too much cost.  Even if some subgraphs of the query graph are matched in the data, we will not attempt to assemble the matches together without following the join order.

The query decomposition approach can still suffer from having to maintain too many partial matches.  If a subgraph of the query graph is highly \textsl{frequent}, we will end up tracking a large number of partial matches corresponding to that subgraph.  Unless we have quantitative knowledge about how these partial matches transition into larger matches, we face the risk of tracking a large number of non-promising matching subgraphs.  The ``Lazy Search" approach outlined earlier in the introduction enhances this further.  For any new edge, we search for a query subgraph if and only if it is the most selective subgraph in the query or if one of the either vertices in that edge participates in a match with the preceding (query) subgraph in the join order.

This section is dedicated towards introducing the data structures and algorithms for dynamic graph search.  We begin with introducing the SJ-Tree structure (section \ref{subsubsec:join_tree_properties}) and then proceed to present the basic algorithms (Algorithm 1 and 2).   The ``Lazy Search"-enhanced version is introduced later in section \ref{sec:Lazy Search}.  Automated generation of SJ-Tree is covered in section \ref{sec:Query Optimization}.

\label{sec:Incremental Query Processing}


\subsection{Subgraph Join Tree (SJ-Tree)}
\label{subsubsec:join_tree_properties}
We introduce a tree structure called \emph{Subgraph Join Tree (SJ-Tree)}.   SJ-Tree defines the decomposition of the query graph into smaller subgraphs and is responsible for storing the partial matches to the query.  Figure \ref{fig:join_tree} shows the decomposition of an example query.  Each of the rectangular boxes with dotted lines will be represented as a node in the SJ-Tree.  The query subgraphs shown inside each ``box" will be stored as a node property described below.  \\

\textsc{Definition \ref{subsubsec:join_tree_properties}.1 } A SJ-Tree $T$ is defined as a binary tree comprised of the node set $N_T$.  Each $n \in N_{T}$ corresponds to a subgraph of the query graph $G_q$.   Let's assume $V_{SG}$ is the set of corresponding subgraphs and $|V_{SG}| = |N_T|$.  Additional properties of the SJ-Tree are defined below.

\textsc{Definition \ref{subsubsec:join_tree_properties}.2 } A \textsl{Match} or a \textsl{Partial Match} is as a set of edge pairs.  Each edge pair represents a mapping between an edge in a query graph and its corresponding edge in the data graph.

\textsc{Definition \ref{subsubsec:join_tree_properties}.3} Given two graphs $G_1 = (V_1, E_1)$ and $G_2 = (V_2, E_2)$, the join operation is defined as $G_3 = G_1 \Join G_2$, such that $G_3 = (V_3, E_3)$ where $V_3 = V_1 \cup V_2$ and $E_3 = E_1 \cup E_2$. \\


\textsc{Property 1.} The subgraph corresponding to the root of the SJ-Tree is isomorphic to the query graph.  Thus, for $n_r = root\lbrace T\rbrace$, $V_{SG}\lbrace n_r \rbrace \equiv G_q$.

\textsc{Property 2.} The subgraph corresponding to any internal node of $T$ is isomorphic to the output of the join operation between the subgraphs corresponding to its children.  If $n_l$ and $n_r$ are the left and right child of $n$, then $V_{SG}\lbrace n \rbrace = V_{SG}\lbrace n_l \rbrace \Join V_{SG}\lbrace n_r \rbrace$.

Therefore, each leaf of the SJ-Tree represent subgraphs that we want to search for (perform subgraph isomorphism) on the streaming updates.  Internal nodes in the SJ-Tree represents subgraphs that result from the joining of subgraphs returned by the subgraph isomorphism operations.

\textsc{Property 3.} Each node in the SJ-Tree maintains a set of \textsl{matches}.  We define a function $matches(n)$ that for any node $n \in N_T$, returns a set of subgraphs of the data graph.  If $M = matches(n)$, then $\forall G_m \in M$,  $G_m \equiv V_{SG}\lbrace n \rbrace$.

\textsc{Property 4.} Each internal node $n$ in the SJ-Tree maintains a subgraph, CUT-SUBGRAPH($n$) that
equals the \textit{intersection} of the query subgraphs of its child nodes.  \\


For any internal node $n \in N_T$ such that CUT-SUBGRAPH$(n) \neq \emptyset$, we also define a \textit{projection operator} $\Pi$.  Assume that $G_1$ and $G_2$ are isomorphic, $G_1 \equiv G_2$.  Also define $\Phi_V$ and $\Phi_E$ as functions that define the bijective mapping between the vertices and edges of $G_1$ and $G_2$.  Consider $g_1$, a subgraph of $G_1$: $g_1 \subseteq G_1$.  Then $g_2 = \Pi(G_2, g_1)$ is a subgraph of $G_2$ such that $V(g_2) = \Phi_V(V\left(g_1\right))$ and $E(g_2) = \Phi_E(E\left(g_1\right))$.


Our decision to use a binary tree as opposed to an n-ary tree is influenced by the simplicity and lowering the combinatorial cost of joining matches from multiple children.  With the properties of the SJ-Tree defined, we are now ready to describe the graph search algorithm.

\subsection{Dynamic Graph Search Algorithm}
\label{sec:Continuous Query Algorithm}


\begin{algorithm}
\caption{DYNAMIC-GRAPH-SEARCH($G_d$, T, edges)}
\label{algo:process_cont_query}
\begin{algorithmic}[1]
\State $leaf$-$nodes = $GET-LEAF-NODES$(T)$
\ForAll {$e_s \in edges$}
\State UPDATE-GRAPH($G_d, e_s$)
\ForAll {$n \in leaf$-$nodes$}
\State $g^q_{sub} = $GET-QUERY-SUBGRAPH$(T, n)$
\State $matches = $SUBGRAPH-ISO($G_d, g^q_{sub}, e_s$)
\If {$matches \neq \emptyset$}
\ForAll {$m \in matches$}
\State UPDATE-SJ-TREE$(T, n, m)$
\EndFor
\EndIf
\EndFor
\EndFor
\end{algorithmic}
\end{algorithm}

We begin with describing our dynamic graph search algorithm (Algorithm \ref{algo:process_cont_query} and \ref{algo:update_match_tree}).  The input to DYNAMIC-GRAPH-SEARCH is the dynamic graph so far $G_d$, the SJ-Tree ($T$) corresponding to the query graph and the set of incoming edges.  Every incoming edge is first added to the graph (Algorithm 1, line 3).  Next, we iterate over all the query subgraphs to search for matches containing the new edge (line 5-6).  Any discovered match is added to the SJ-Tree (line 9).

Next, we describe the UPDATE-SJ-TREE function.  Each node in the SJ-Tree maintains its sibling and parent node information (Algorithm 2, line 1-2).  Also, each node in the SJ-Tree maintains a hash table (referred by the \textit{match-tables} property in Algorithm 2, line 4).  GET() and ADD() provides lookup and update operations on the hash tables.  Each entry in the hash table refers to a \textsl{Match}.   Whenever a new matching subgraph $g$ is added to a node in the SJ-Tree, we compute a key using its projection $(\Pi(g))$ and insert the key and the matching subgraph into the corresponding hash table (line 12).  When a new match is inserted into a leaf node we check to see if it can be combined (referred as JOIN()) with any matches that are contained in the collection maintained at its sibling node.  A successful combination of matching subgraphs between the leaf and its sibling node leads to the insertion of a larger match at the parent node.  This process is repeated recursively (line 11) as long as larger matching subgraphs can be produced by moving up in the SJ-Tree.  A complete match is found when two matches belonging to the children of the root node are combined successfully.

\textsc{Example} Let us revisit Figure \ref{fig:join_tree} for an example.  Assuming we find a match with the query subgraph  containing a single ``friend" edge (e.g. $\lbrace$(``George", ``friend", ``Sutanay")$\rbrace$), we will probe the hash table in the leaf node with ``likes" edges.  If the hash table stored a subgraph such as $\lbrace$(``Sutanay", ``likes", ``Santana")$\rbrace$, the JOIN() will produce a 2-edge subgraph $\lbrace$(``George", ``friend", ``Sutanay"), (``Sutanay", ``likes", ``Santana")$\rbrace$.  Next, it will be inserted into the parent node with 2-edges.  The same process will be subsequently repeated, beginning with the probing of the hash table storing matches with subgraphs with a ``follows" edge.

\begin{algorithm}
\caption{UPDATE-SJ-TREE($node, m)$}
\label{algo:update_match_tree}
\begin{algorithmic}[1]
\State $sibling = sibling[node]$
\State $parent = parent[node]$
\State $k = $GET-JOIN-KEY(CUT-SUBGRAPH[$parent$], $m$)
\State $H_s$ = match-tables[$sibling$]
\State $M^k_s$ = GET($H_s, k$)
\ForAll {$m_s \in M^k_s$}
\State $m_{sup}$ = JOIN($m_s, m$)
\If {parent = root}
\State PRINT('MATCH FOUND :  ', $m_{sup}$)
\Else
\State UPDATE-SJ-TREE($parent, m_{sup}$)
\EndIf
\EndFor
\State ADD(match-tables$[node], k, m$)
\end{algorithmic}
\end{algorithm}


\section{Analysis of Dynamic Graph Search Algorithm}
\label{sec:Continuous Query Algorithm}

At this point, it is probably obvious that different SJ-Tree structures can be generated from the same query graph.  Later in the paper, we provide example query decompositions in Figure \ref{fig:new_figs/apps/netflow_decomp}.  While multiple factors can lead to generation of different SJ-Trees, one primary factor is our choice for granularity of decomposition, the size and the structure of the subgraphs we decompose the query to.  As we will establish through extensive analysis through this paper, there is value in establishing a standard set of small subgraphs that are efficient to search for in a real-time setting.  Henceforth, we often refer to these set of small subgraphs as \textsl{search primitives} or simply \textsl{primitives}.  As a first step to understand the speed-memory tradeoff associated with different choices for primitives, we begin with the complexity analysis of the dynamic graph search described in Algorithm \ref{algo:process_cont_query} and \ref{algo:update_match_tree}.  A key operation in Algorithm \ref{algo:process_cont_query} is the process of subgraph isomorphism around every new edge in the graph.  Therefore, we exclusively focus on the complexity analysis in terms of 1-3 edge subgraphs as candidates for search primitives.

\textsc{Single Edge Subgraphs}  When the query graph ($g^q_{sub}$ in Algorithm 1, line 5) contains a single edge, checking if an edge from the data graph ($e_s$) matches the query edge require comparing the types and potentially other attributes of the edges.  Depending on the query constraint, we may need to look up the node label to perform a string comparison or evaluate a regular expression.  The node labels or any other node-specific properties are stored in an array leading to constant time access to node labels.  Therefore, a single-edge query can be matched in $O(1)$ time.

\textsc{Triads} Assume that the query graph is a triad with three vertices $v_1$, $v_2$ and $v_3$, and edges ordered as $e_1=(v_1, v_2), e_2=(v_2, v_3), e_3=(v_3, v_1)$.  For any edge $e$ in the data graph, we can detect a match with $e_1$ in constant time.  If $e$ is matched, we search the neighborhood of the vertex that matches with $v_2$ to search for $e_2$.  Denoting this vertex as $v^{'}_2$, the cost of this second level of search is $O(degree(v^{'}_2))$.  In case of a 3-edge subgraph, each of the successful second level searches proceed to find a match for the third edge.  Thus, the cost of a 2-edge subgraph is $O(degree(v^{'}_2))$ and a 3-edge subgraph is $O(degree(v^{'}_2)*degree(v^{'}_3))$.  We can refine these estimates to obtain an average cost of the search as $O(\bar{d_2})$ for a 2-edge subgraph and $O(\bar{d_2}\bar{d_3})$ for a 3-edge subgraph, where $\bar{d_2}$ and $\bar{d_3}$ are the average degree of the vertices in the graph for the types of $v_2$ and $v_3$.

The next step is to estimate a cost for the SJ-Tree update operation (Algorithm \ref{algo:update_match_tree}).  We begin with the hash-join operation (Algorithm \ref{algo:update_match_tree}, line 7).

Assume the frequency of a graph $g^i_q$ is $n_i$, where the frequency of a subgraph is defined as the count of its instances over an edge stream of length $N$.  Therefore, over $N$ edges, we can expect $O(n_1)$ matches for $g^1_q$ and $O(n_2)$ matches for $g^2_q$.  Therefore, $H_2$ (hash table associated with the SJ-Tree node representing $g^2_q$) will be probed for a match $O(n_1)$ times over $N$ edges and $H_1$ (associated with the SJ-Tree node representing $g^1_q$) will be probed $O(n_2)$ times within the same period.

If we knew the frequency of $G_q$, henceforth referred as $f_S(G_q)$, then we can also estimate the number of new subgraphs that will be produced as the result of the hash-joins.  Given that the frequency of the larger subgraph can not exceed that of the more selective component we can approximate $O(n(G^q)) \simeq min\left(O(n_1), O(n_2))\right)$.  Therefore, the average work for every incoming edge in the graph can be expressed as, \\ $\left(f_S(g^1_q) + f_S(g^2_q) + O(n_1) + O(n_2) + min\left(O(n_1), O(n_2))\right)\right)/N$.

The Hash-Join combined with leaf level searches provides the simplest example of a SJ-Tree, a binary tree with height 1.  In this section, we analyze the time complexity of the query processing as it happens in a multi-level SJ-Tree.  Given any non-leaf node $n$, we can obtain the expression for average work by adapting the complexity expression shown above.  Note that if a child of $n$, denoted by $n_c$, is not a leaf level node but an internal node, then the term corresponding to the search cost ($f_S(g)$) disappears.  Additionally, we can replace the search cost with the cost corresponding to the average work incurred by the subtree rooted by $n_c$.  Therefore, given a SJ-Tree ($T_{sj}$) the average work ($C(T_{sj})$) can be obtained by recursive computation from the root.  $C(T_{sj}) = C(root(T_{SJ}))$

\section{Lazy Search}
\label{sec:Lazy Search}

Revisiting our example from Figure \ref{fig:join_tree}, it is reasonable to assume that the ``friend" relation is highly frequent in the data.  If we decomposed the query graph all the way to single edges then we will be tracking all edges that match ``friend".  Clearly, this is wasteful.  One may suggest decomposing the query to larger subgraphs.  However, it will also increase the average time incurred in performing subgraph isomorphism.  Deciding the right granularity of decomposition requires significant knowledge about the dynamic graph.  This motivates us to introduce a new algorithmic extension.

Assume the query graph $G_q$ is partitioned into two subgraphs $g_1$ and $G^1_q$.  We use the notation $G^k_q$ to indicate what remains of $G_q$ after the $k$-th iteration in the decomposition process.  If the probability of finding a match for $g_1$ is less than the probability of finding a match for $G^1_q$, then it is always desirable to search for $g_1$ and look for $G^1_q$ only where an occurrence of $g_1$ is found.  Therefore, we select $g_1$ to be the most selective edge or 2-edge subgraph in the query graph and always search for $g_1$ around every new edge in the graph.  Once we detect subgraphs in $G_d$ that match with $g_1$, we follow the same approach to search for $G_q$ in their neighborhood.  We partition $G^1_q$ further into two subgraphs: $g_2$ and $G^2_q$, where $g_2$ is another 1-edge or 2-edge subgraph.

\textsc{Data Structures}  With the SJ-Tree, the partitioning of $G_q$ is done upfront at the query compile time with $g_1$, $g_2$ etc becoming the leaves of the tree.  The main difference between Lazy Search and that of Algorithm 2 is that we will be searching for $g_2$ only around the edges in $G_d$ where a match with $g_1$ is found.  Therefore, for every vertex $u$ in $G_d$, we need to keep track of the $g_i$-s such that $u$ is present in the matching subgraph for $g_i$.  We use a bitmap structure $M_b$ to maintain this information.  Each row in the bitmap refers to a vertex in $G_d$ and the $i$-th column refers to $g_i$, or the $i$-th leaf in the SJ-Tree.  If the search for subgraph $g_i$ is enabled for vertex $u$ in $G_d$, then $M_b[u][i] = 1$ and zero otherwise.  Whenever a matching subgraph $g'$ for $g_i$ is discovered, we turn on the search for $g_{i+1}$ for all vertices in $V(g')$.  This is accomplished by setting $M_b[v][i+1] = 1$ where $v \in V(g')$.


\textsc{Robustness with Subgraph Arrival Order} Consider a SJ-Tree with just two leaves representing query subgraphs $g_1$ and $g_2$, with $g_1$ representing the \textit{more selective} left leaf.  The above strategy is not robust to the arrival order of matches.  Assume $g^{'}_1$ and $g^{'}_2$ are subgraphs of $G_d$ that are isomorphic to $g_1$ and $g_2$ respectively.  Together, $g^{'}_1 \times g^{'}_2$ is isomorphic to the query graph $G_q$.  Because we are searching for $g_1$ on every incoming edge, $g^{'}_1$ will be detected as soon as it appears in the data graph.  However, we will detect $g^{'}_2$ only if appears in $G_d$ after $g^{'}_1$.  If $g^{'}_2$ appeared in $G_d$ before $g^{'}_1$ we will not find it because we are not searching for $g_2$ all the time.

We introduce a small change to address this temporal ordering issue.  Whenever we enable the search on a node in the data graph, we also perform a subgraph search around the node to find any match that has occurred earlier.  Thus, when we find $g_1$ and enable the search for $g_2$ on every subsequent edge arrival, we also perform a search in $G_d$ looking for $g_1$.  This ensures that we will find $g_2$ even if it appeared before $g_1$.

Algorithm \ref{algo:lazy_search} summarizes the entire process.  Lines 2-3 loop over all news edges arriving in the graph and update the graph.  Next, given a new edge $e_s$, for each node in the SJ-Tree, we check to see if we should be searching for its corresponding subgraph around $e_s$ (lines 4-8).  The DISABLED() function queries the bitmap index and returns \textit{true} if the corresponding search task is disabled.  GET-QUERY-SUBGRAPH returns the query subgraph $g^q_{sub}$ corresponding to node $n$ in the SJ-Tree (line 9).  Next, we search for $g^q_{sub}$ using a subgraph-isomorphism routine that only searches for matches containing at least one of the end-point vertices of $e_s$ ($u$ and $v$, mentioned in line 5-6).   For each matching subgraph found containing $u$ or $v$, we enable the search for the query subgraph corresponding the sibling of $n$ in the SJ-Tree.  If $n$ was not left-deep most node in the SJ-Tree, then we also query the left sibling to probe for potential join candidates (QUERY-SIBLING-JOIN(), line 16).  Any resultant joins are pushed into the parent node and the entire process is recursively repeated at one level higher in the SJ-Tree.

\begin{algorithm}
\caption{LAZY-SEARCH($G_d$, T, edges)}
\label{algo:lazy_search}
\begin{algorithmic}[1]
\State $leaf$-$nodes = $GET-LEAF-NODES$(T)$
\ForAll {$e_s \in edges$}
\State UPDATE-GRAPH($G_d, e_s$)
\ForAll {$n \in leaf$-$nodes$}
\State $u = $src$(e_s)$
\State $v = $dst$(e_s)$
\If {DISABLED(u, n) AND DISABLED(v, n)}
\State continue
\EndIf
\State $g^q_{sub} = $GET-QUERY-SUBGRAPH$(T, n)$
\State $matches = $SUBGRAPH-ISO($G_d, g^q_{sub}, e$)
\ForAll {$m \in matches$}
\If {$n = 0$}
\State ENABLE-SEARCH-SIBLING$(n, m)$
\Else
\State $M_j$ = QUERY-SIBLING-JOIN($n, m$)
\State $p$ = PARENT($n$)
\ForAll {$m_j \in M_j$}
\State UPDATE($p, m_j$)
\State ENABLE-SEARCH-SIBLING$(p, m)$
\EndFor
\EndIf
\EndFor
\EndFor
\EndFor
\end{algorithmic}
\end{algorithm}
\section{SJ-Tree Generation}
\label{sec:Query Optimization}

Here we address the topic of automatic generation of the SJ-Tree from a specified query graph.  We begin with introducing key definitions, followed by the decomposition algorithm.

\textsc{definition} \textbf{Subgraph Selectivity} Given a large typed, directed graph $G$, the selectivity of a typed, directed subgraph $g$ with $k$-edges (denoted as $S(g)$) is the ratio of the number of occurrences of $g$ and the total number of all $k$-edge subgraphs in $G$.  Instances of $g$ may overlap with each other.

\textsc{definition} \textbf{Selectivity Distribution} The selectivity distribution of a set of subgraphs $G_k$ is a vector containing the selectivity for every subgraph in $G_k$.  The subgraphs are ordered by their frequencies in ascending order.




We present a greedy algorithm (Algorithm \ref{algo:build_sj_tree})  for decomposing a query graph into its subgraphs and generating a SJ-Tree. Our choice for the greedy heuristic is motivated by extensive survey of the literature on optimal join order determination in relational databases \cite{krishnamurthy1986optimization, wu2003structural, hellerstein1993predicate}.  A key conclusion of the survey states that \textsl{left-deep join plans} (or left deep binary trees in this case) is one of the best performing heuristics.  The above mentioned studies point to a large body of research using techniques such as dynamic programming and genetic algorithms to find the optimal join order.  Nonetheless, finding the lowest cost join order or using a cost-driven join order determination remains an interesting problem in graph databases, and the approaches based on minimum spanning trees or approximate vertex cover can provide an initial path forward.

Inputs to  Algorithm \ref{algo:build_sj_tree} are the query graph $G_q$ and an ordered set of primitives $M$.  Our goal is to decompose $G_q$ into a collection of (possibly repeated) subgraphs chosen from $M$.  Entries of $M$ are sorted in ascending order of their subgraph selectivity.  Given a query graph $G_q$, the algorithm begins with finding the subgraph with the lowest selectivity in $M$.  This subgraph is next removed from the query graph and the nodes of the removed subgraph are pushed into a ``frontier" set.  We proceed by searching for the next selective subgraph that includes at least one node from the frontier set.  We continue this process until the query graph is empty.   SUBGRAPH-ISO performs a subgraph isomorphism operation to find an instance of $g_M$ in $G_q$.  Algorithm \ref{algo:build_sj_tree} uses two versions of SUBGRAPH-ISO.  The first version uses three arguments, where the second argument is a vertex id $v$.  This version of SUBGRAPH-ISO searches $G_q$ for instances of $g_M$ by only searching in the neighborhood of $v$.  The other version accepting two arguments searches entire $G_q$ for an instance of $g_M$.  REMOVE-SUBGRAPH accepts two graphs as argument, where the second argument ($g_{sub}$) is a subgraph of the first graph ($G_q$).  It removes all edges in $G_q$ that belong to $g_{sub}$.  A vertex is removed from $G_q$ only when the edge removal results in a disconnected vertex.

\begin{algorithm}
\caption{BUILD-SJ-TREE$(G_q, M)$}
\label{algo:build_sj_tree}
\begin{algorithmic}[1]
\State $frontier = \emptyset$
\While {$|V(G_q)| > 0$}
\State $g_{sub} = \emptyset$
\ForAll {$g_M \in M$}
\If {$frontier \neq \emptyset$}
\ForAll {$v \in frontier$}
\State $g_{sub} = $SUBGRAPH-ISO$(G_q, v, g_M)$
\State break
\EndFor
\Else
\State $g_{sub} = $SUBGRAPH-ISO$(G_q, g_M)$
\EndIf
\EndFor
\If {$g_{sub} \neq \emptyset$}
\State $frontier = frontier \cup V(g_{sub})$
\State $G_q = $REMOVE-SUBGRAPH$(G_q, g_{sub})$
\EndIf
\EndWhile
\end{algorithmic}
\end{algorithm}

\subsection{Selectivity Estimation of Primitives}

We propose computing the selectivity distribution of primitives by processing an initial set of edges from the graph stream.  For experimentation purposes we assume that the selectivity order remains the same for the dynamic graph when we perform the query processing.  This work does not focus on modeling the accuracy of this estimation.  Modeling the impact on performance when the actual selectivity order deviates from the estimated selectivity order is an area of ongoing work.

Which subgraphs are good candidates as entries of $M$?  Following are two desirable properties for entries in $M$:  1) the cost for subgraph isomorphism should be low.  2) Selectivity estimation of these subgraphs should be efficient as we will need to periodically recompute the estimates from a graph stream.   Based on these two criteria, we select single edge subgraphs and 2-edge paths as primitives in this study.  Computing the selectivity distribution for single-edge subgraphs resolves to computing a histogram of various edge types.  The selectivity distribution for 2-edge paths on a graph with $V$ nodes, $E$ vertices and $k$ unique edge types can be done in $O(V(E+k^2))$ time.  Algorithm \ref{algo:sample_paths} provides a simple algorithm to count all 2-edge paths.  In our experiments, computing the path statistics for a network traffic dataset with 800K nodes and nearly 130 million edges takes about 50 seconds without any code optimization.

Algorithm \ref{algo:sample_paths} uses a Counter() data structure, which is a hash-table where given a key, the corresponding value indicates the number of times the key occurred in the data.  A Counter() is updated via the UPDATE routine, which accepts the counter object, a key value and an integer to increment the corresponding key count.    We iterate over all vertices in the input graph ($G_d$) (line 2).  For an given vertex $v$, we count the number of occurrences of each unique edge type associated with it (accounting for edge directions).  Line 8 iterates over all unique edge types associated with $v$.  Next, given an edge type $e_1$ and its count $n_1$, we count the number of combinations possible with two edges of same type ($n \choose 2$).  Next, we compute the number of 2-edge paths that can be generated with $e_1$ and any other edge type $e_2$.  We impose the LEXICALLY-GREATER constraint to ensure each edge is factored in only once in the 2-edge path distribution.

Note that we use a $Map()$ function instead of simply using the type associated with every edge.  Most of our target applications have significant  amount edge attributes in the graphs.  As an example, in a network traffic graph we use the protocol information to determine the edge property.  Thus, each network flow with the same protocol (e.g. HTTP, ICMP etc.) are mapped to the same edge type.  Each flow is accompanied by multiple attributes such as source and destination ports, duration of communication etc..  Therefore, we can provide a hash function to map any user defined edge properties to an integer value.  Thus, for queries with constraints on vertex and edge properties, a generic map function factors in both structural and semantic characteristics of the graph stream.

Counting the frequency for larger subgraphs is important.  Given a query graph with $M$ edges, ideally we would like to know the frequency of all subgraphs with size $1, 2, .., M-1$.  Collecting the frequency of larger subgraphs, specifically triangles have received a significant attention in the database and data mining community.  Exhaustive enumeration of all the triangles can be expensive, specially in the presence of high degree vertices in the data.  Approximate triangle counting via sampling for streaming and semi-streaming has been extensively studied in the recent years  \cite{seshadhri2013fast, jha2013space}.  We foresee incorporation of such algorithms to support better query optimization capabilities for queries with triangles.

\begin{algorithm}
\caption{COUNT-2-EDGE-PATHS($G_d$)}
\label{algo:sample_paths}
\begin{algorithmic}[1]
\State $P = Counter()$
\ForAll {$v \in V(G_d)$}
\State $C_v$ = $Counter()$
\ForAll {$e \in Neighbors(G_d, v)$}
\State $e_t = Map(e)$
\State $Update(C_v, e_t, 1)$
\EndFor
\State $E_t = Keys(C_v)$
\ForAll {$e_1 \in E_t$}
\State $n_1 = Count(C_v, e_1)$
\State $key = (e_1, e_1)$
\State $Update(P, key, n_1(n_1-1)/2)$
\ForAll {$e_2 \in $LEXICALLY-GREATER$(E_t, e_1)$}
\State $n_2 = Count(C_v, e_2)$
\State $key = (e_1, e_2)$
\State $Update(P, key, n_1n_2)$
\EndFor
\EndFor
\EndFor
\end{algorithmic}
\end{algorithm}

\subsection{Query Decomposition Strategies}

Algorithm \ref{algo:build_sj_tree} shows that we can generate multiple SJ-Trees for the same $G_q$ by selecting different primitive sets for $M$.  We can initiate $M$ with only 1-edge subgraphs, only 2-edge subgraphs or a mix of both.  As an example, for a 4-edge query graph, the removal of the first 2-edge subgraph can leave us with 2 isolated edges in $G_q$.  At that stage, we will create two leaf nodes in the SJ-Tree with 1-edge subgraphs.  For brevity we refer to both the second and third choice as 2-edge decomposition in the remaining discussions.  Clearly, these 1 or 2-edge based decomposition strategies has different performance implications.  Searching for 1-edge subgraphs is extremely fast.  However, we stand to pay the price with memory usage if these 1-edge subgraphs are highly frequent.  On the contrary, we expect 2-edge subgraphs to be more discriminative.  Thus, we will trade off lowering the memory usage by spending more time searching for larger, discriminative subgraphs on every incoming edge.

\textsc{definition} \textbf{Expected Selectivity}   We introduce a metric called \textsl{Expected Selectivity}, denoted as $\hat{S(T_k)}$.  Given a SJ-Tree $T_k$, the Expected Selectivity is defined as the product of the selectivities of the leaf-level query subgraphs.

$leaves(T_k)$ returns the set of leaves in a SJ-Tree $T_k$.  Given a node $n$, $V_{SG}(T, n)$ returns the subgraph corresponding to node $n$ in SJ-Tree $T$.  Finally, $S(g)$ is the selectivity of the subgraph $g$ as defined earlier.
\begin{equation}
\hat{S(T_k)} = \prod_{n \in leaves(T_k)}S(V_{SG}(T_k, n))
\end{equation}

\textsc{definition} \textbf{Relative Selectivity}   We introduce a metric called \textsl{Relative Selectivity}, denoted as $\xi(T_k, T_1)$.  Given a 1-edge decomposition $T_1$ and another decomposition $T_k$, we define $\xi(T_k, T_1)$ as follows.

\begin{equation}
\xi(T_k, T_1) = \frac{\hat{S(T_k)}}{\hat{S(T_1)}}
\end{equation}



We conclude the section with discussion on two desirable properties of a greedy SJ-Tree generation strategy.

\textsc{Theorem 1} Given the data graph $G_d$ at any time $t$, assume that the query graph $G_q$ is not guaranteed to be present in $G_d$.  Then initiating the search for $G_q$ by searching for $g_{rare}$ where $g_{rare} \subset G_q$ and $\forall g \subset G_q | |E(g)| = |E(g_{rare})|, frequency(g) > frequency(g_{rare})$ is in optimal strategy.

\textsc{Proof}  The time complexity for searching for a $O(1)$ for a 1-edge subgraph and  $O(\bar{d}_v)$ for a 2-edge subgraph.  Therefore, the runtime cost to search for $g_{rare}$ is same as any other subgraph of $G_q$ with the same number of edges.  However, searching for $g_{rare}$ will require minimum space because it has the minimum frequency amidst all subgraphs with same size.  Therefore, searching for $g_{rare}$ is an optimal strategy.


\begin{figure}[htbp]
\centering
\includegraphics[scale=0.35]{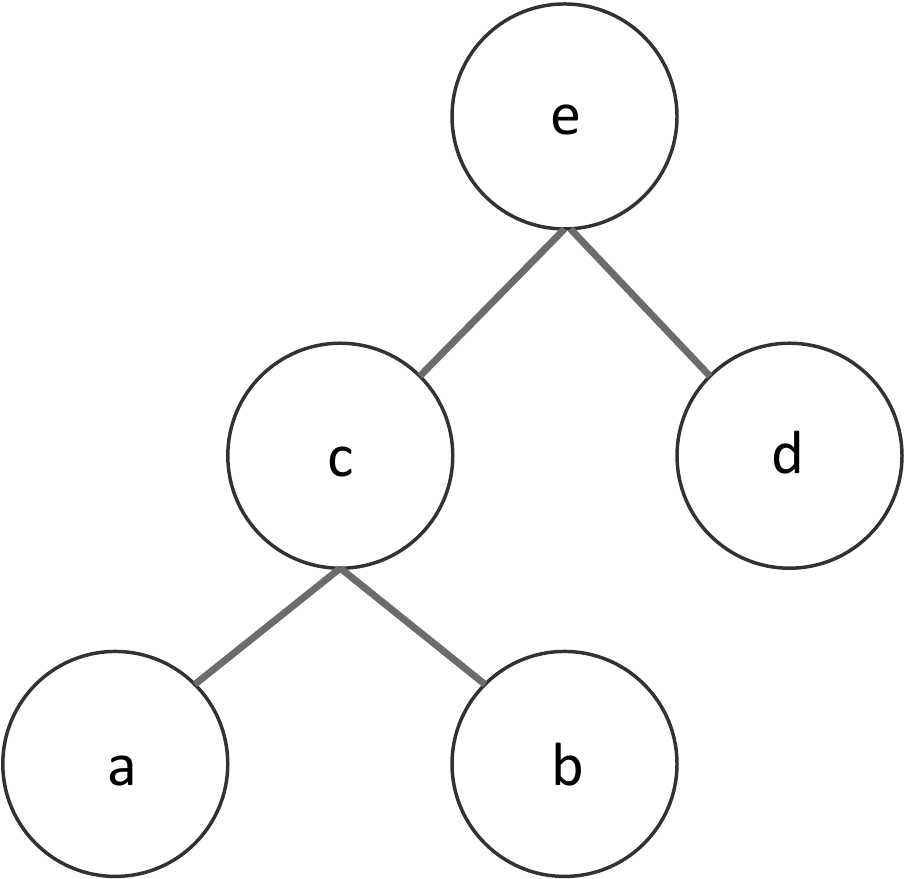}
\caption{}
\label{fig:new_figs/proof_tree}
\end{figure}

\textsc{Theorem 2} Given a set of identical size subgraphs $\lbrace g_k \rbrace$ such that $\cup^n_{k} g_k = G_q$, a SJ-Tree with ordered leaves $g_k \prec g_{k+1}  \prec g_{k+2}$ requires minimal space when $frequency(g_k \Join g_{k+1}) < frequency(g_{k+2})$.


\textsc{Proof} By induction.  Assume a SJ-Tree with three leaves as shown in Figure \ref{fig:new_figs/proof_tree}.  Following the definitions of SJ-Tree, this is a left-deep binary tree with 3 leaves.  Therefore, $frequency(c )$ denoted in shorthand as $f(c )$ $f(c )$ = $min(f(a), f(b))$.  Substituting for the frequency of $c$, space requirement for this tree $S(T) = f(a) + f(b) + f(d) + min(f(a), f(b))$.  Thus, the space requirement for this tree is minimum if $f(a) < f(b) < f(c )$.

Now we can consider any arbitrary tree where $T_{n}$ refers to a tree with a left subtree $T_{n_1}$ and a right child $l_{n+2}$.  Above shows that $T_1$ constructed as above will have minimum space requirement, and so will $T_2$ if  $f(a) < f(b) < f(c ) < f(d)$.


\textsc{Observation 3}  Given $g_k$, a subgraph of query graph $G_q$, it is efficient to decompose $g_k$ if there is a subgraph $g \subset g_k$, such that frequency$(g) > \left(\frac{frequency(g_k)}{\bar{d}|V(g_k)|}\right)$, where $\bar{d}$ is the average vertex degree of the data graph and $|V(g_k)|$ is the number of vertices in $g_k$.

\textsc{Proof}  Given a graph $g$, the average cost for searching for another graph that is larger by a single edge is $\bar{d}$ multiplied by the number of vertices in $g_k$, and the proof follows.




\section{Experimental Studies}
\label{sec:Experimental Results}

We present experimental analysis on two real-world datasets (New York Times \footnote[1]{http://data.nytimes.com} and Internet Backbone Traffic data \footnote[1]{http://www.caida.org}), and a synthetic streaming RDF benchmark.  The experiments are performed to answer questions in the following categories.

\begin{enumerate}
\item \textsc{Studying Selectivity Distribution} What does the selectivity distribution of 2-edge subgraphs look like in real world datasets?  What is the duration of time for which the selectivity distribution or selectivity order of 2-edge subgraphs remains static?
\item \textsc{Comparison between Search strategies}  In the previous sections, we introduced two different choices for query decomposition (1-edge vs 2-edge path based) and two different choices for query execution (lazy vs non-lazy).  How do the strategies compare?
\item \textsc{Automated strategy selection} Given a dynamic graph and a query graph, can we choose an effective strategy using their statistics?
\end{enumerate}

\subsection{Experimental setup}

The experiments were performed on a 32-core Linux system with 2.1 GHz AMD Opteron processors, and with 64 GB memory.  The code was compiled with g++ 4.7.2 compiler with -O3 optimization.

Given a pair of data graph and query graph, we perform either of two tasks: 1) query decomposition and 2) query processing.

\textsl{Query decomposition: } Query decomposition involves loading the data graph, collecting 1-edge and 2-edge subgraph statistics and performing query decomposition using the selectivity distribution of the subgraphs.  The SJ-Tree generated by the query decomposition algorithm is stored as an ASCII file on disk.

\textsl{Query processing: } The query processing step begins with loading the query graph in memory, followed by initialization of the SJ-Tree structure from the corresponding file generated in the query decomposition step.  We initialize the data graph in memory with zero edges.   Next, edges parsed from the raw data file are streamed into the data graph.  The continuous query algorithm is invoked after each AddEdge() call to the data graph.



\subsection{Data source description}

Summaries of various datasets used in the experiments are provided in Table 1.  We tested each dataset  with a set of randomly generated queries.  The following describes the individual datasets and test query generation.

\begin{table*}[htbp]
\caption{Summary of test datasets}
\begin{center}
\begin{tabular}{|c|c|c|c|}
\hline \hline
Dataset & Type & Vertices & Edges \\
\hline
New York Times & Online News & 64,639 & 157,019 \\
\hline
Internet Backbone Traffic & Network traffic & 2,491,915 & 19,550,863 \\
\hline
LSBench/CSPARQL Benchmark & RDF Stream & 5,210,099 & 23,320,426 \\
\hline
\end{tabular}
\end{center}
\label{default}
\end{table*}

\textbf{New York Times}:   The New York Times dataset contains articles collected from 2013 July-September time period using Version 3 of its data collection API available at \url{data.nytimes.com}.  Each article in the dataset contains a number of facets that belong to four types of entities: person, geo-location, organization and topic.  Each of the articles and facets are represented as vertices in the graph.  Each edge that connects an article with a facet carries a timestamp that is the publication time of the article.  The New York Times dataset was tested with a set of 10 randomly generated k-partite graphs.

\textbf{Network Traffic} The second dataset is an internet backbone traffic dataset obtained from \url{www.caida.org}.  CAIDA (Cooperative Association for Internet Data Analysis) is a collaborative program that provides a wide collection of network traffic data.   We used the ``CAIDA Internet Anonymized Traces 2013 Dataset" for experimentation.  The dataset contains 22 million network traffic flow (subsequently referred to as \textsl{netflow}) records collected over a \emph{one minute period}.  We excluded the traffic to/from IP addresses matching patterns 10.x.x.x or 192.168.x.x.  These address spaces refer to private subnets and a communication from a given IP address from these spaces can actually refer to multiple physical hosts in the real word.  As an example, every internet service provider configures the routers or machines inside a home network with IPs selected from the private IP address range.  Therefore, if we see a request from 192.168.1.1 to google.com, there is no way to determine the exact origin of this communication.  From a graph perspective, allowing private IP address and the subsequent aggregation of communication will result in the creation of vertices with giant neighbor lists, which will surely impact the search performance.  A detailed list of use cases describing subgraph queries for cyber traffic monitoring are described in   \cite{Joslyn:2013:MSC:2484425.2484428}.

\textbf{Social Media Stream} Our final test dataset is a synthetic RDF social media stream available from the Linked Stream Benchmark (LSBench) \cite{DBLP:journals/ijsc/BarbieriBCVG10}.  We generated the dataset using the sibgenerator utility with 1 million users specified as the input parameter.  The generated graph has a static and a streaming component.  The static component refers to the social network with user profiles and social network relationships.  The streaming component includes 3 streams.  The \textit{GPS stream} includes user checkins at various locations.  The \textit{Post and Comments stream} includes posts and comments by the users, subscriptions by users to forums, and a stream of ``likes" and ``tags".  Finally, the \textit{photo stream} includes information about photos uploaded by users, and ``tags" and ``likes" as applied to photos.


\begin{figure*}[htbp]
\centering
\subfigure[Online news - New York Times]{\includegraphics[scale=0.3]{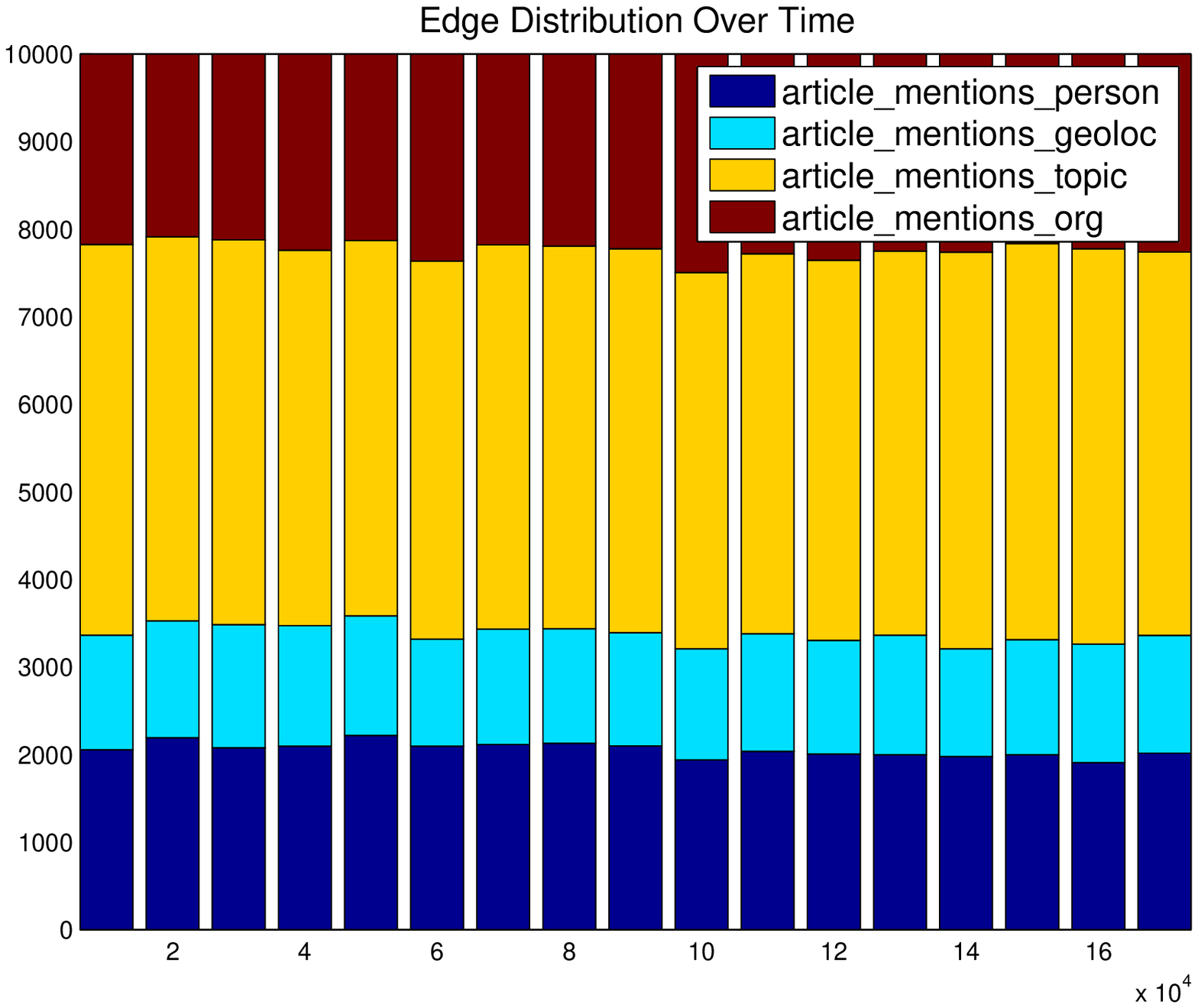}}\quad
\subfigure[Internet Backbone Traffic - CAIDA]{\includegraphics[scale=0.3]{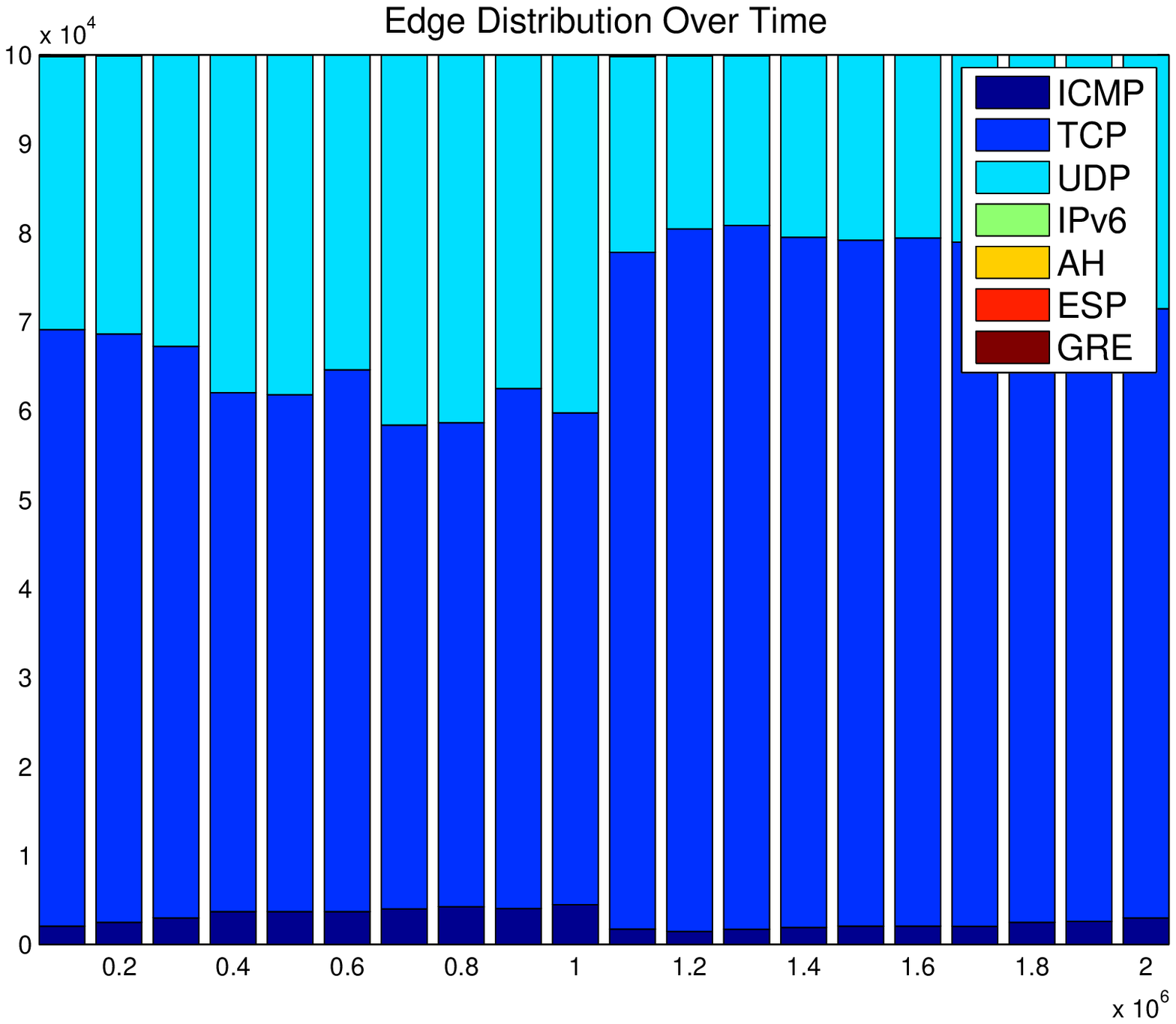}}\quad
\subfigure[Synthetic social data stream in RDF]{\includegraphics[scale=0.3]{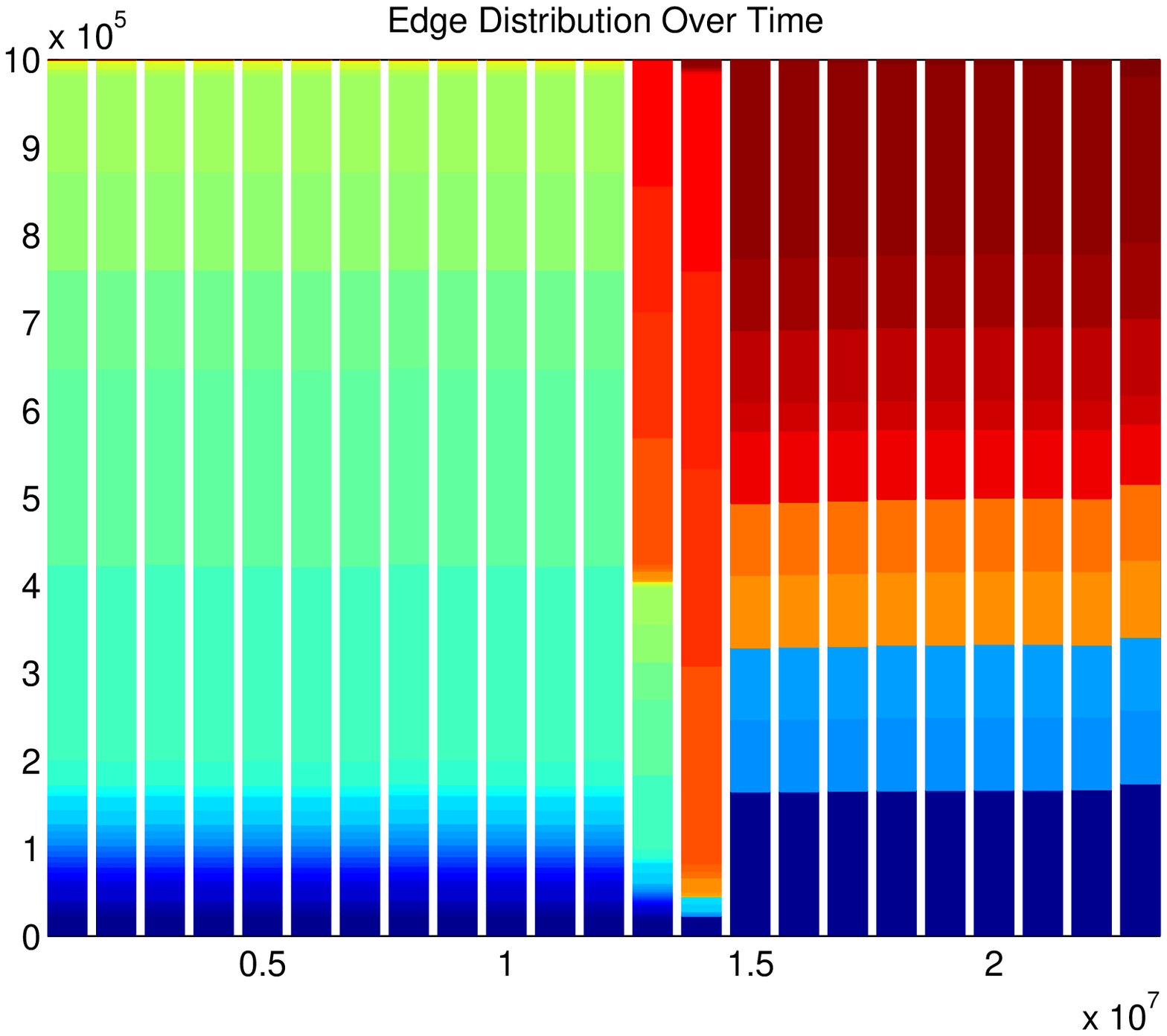}}
\caption{Edge type distribution shown with the evolution of the dynamic graph.}
\label{fig:edge_distribution}
\end{figure*}

\subsection{Selectivity Distribution}
Figure \ref{fig:edge_distribution} shows the edge distribution plotted over time.  X-axis shows the number of cumulative edges in the graph as it is growing.  The plotted distribution is not cumulative.  The edge distribution is collected after fixed intervals.  The interval is 10 thousand, 100 thousand and 1 million respectively.  There are 4, 7, and 45 edge types in these datasets.  The first half of the RDF dataset contains data for a simulated social network.  The second half contains simulated data about the activities in the network such as posts, and checkins at locations  The shift in the edge distribution around the mid point reflects these different characteristics.  The key observation is that the relative order of different types of edges stays similar even as the graph evolves.

\begin{figure}[htbp]
\centering
\subfigure[Synthetic social data stream in RDF]{\includegraphics[scale=0.3]{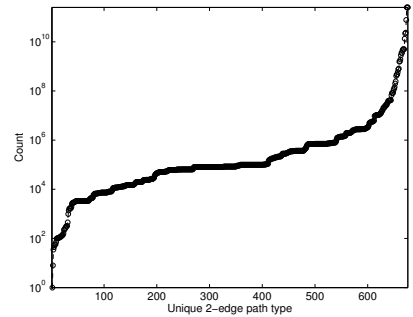}}
\caption{2-edge path distribution in each test data set.  Each point on X-axis represents a unique 2-edge path and Y-axis shows its corresponding count.}
\label{fig:path_distribution}
\end{figure}

There were 14, 62 and 676 unique 2-edge paths present in the New York Times, netflow and LSBench datasets.  Figure \ref{fig:path_distribution} shows the 2-edge path distribution for the LSBench dataset.  We found a small number of 2-edge subgraphs to dominate the distribution across all the datasets.  Other datasets show a similarly skewed distribution, and was omitted for space.  The skew is heaviest for the LSBench dataset, which is expected given the higher number of unique edge types and the larger size of the dataset.

The goal of this analysis was to observe the variability in the selectivity distribution over time.  The selectivity distribution is expected to vary over time.  However, it is the relative order of the unique single edge or 2-edge subgraphs that matters from the query decomposition perspective.  For each of the test datasets, we took multiple snapshots of the selectivity order and found it to be stable, except with fluctuations for the very low frequency components (data points on the left end of the distributions in Fig. \ref{fig:path_distribution}).  Significant changes in the selectivity order can adversely impact the performance of the query.  Estimating the duration over which the selectivity ordering stays stable for a given data stream, quantification of errors based on shift in the distribution, and adapting the query algorithm to handle such shifts is reserved for future work.

\subsection{Query Performance Analysis}

This section presents query performance results obtained through query sweeps on each of the three datasets.  For each query, we collect performance from 4 different query execution strategies obtained by 1-edge or 2-edge decomposition of a query graph and the lazy vs. track everything approach adapted by the query algorithm.  The following tags are used to describe the plots in the remainder of the paper: a) ``\textit{Single}": 1-edge decomposition, search tracks all matching subgraphs in SJ-tree, b) ``\textit{SingleLazy}": 1-edge based query decomposition, use ``Lazy" approach to search, c) ``\textit{Path}": 2-edge decomposition, search tracks all matching subgraphs in SJ-Tree, and d) ``\textit{PathLazy}": 2-edge decomposition with ``Lazy" search.

\subsubsection{New York Times}

\begin{figure}[h]
\centering
\includegraphics[scale=0.45]{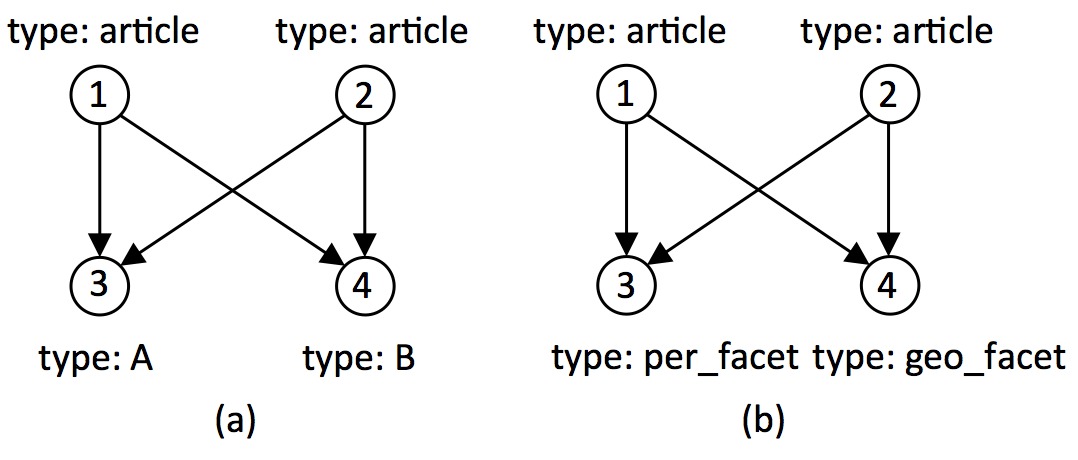}
\caption{Query template for New York Times.}
\label{fig:nyt_template}
\end{figure}

We begin with our analysis on New York Times (NYT) data.  Ten query graphs were generated using the template as shown in Figure \ref{fig:nyt_template}.  Type of vertices 1 and 2 were kept fixed as ``articles", whereas the type of vertices 3 and 4 were permuted between author, location, organization and topic.   The edge types were also changed to keep them consistent with the types of end-point vertices.  For example, setting vertex 3 as ``author" from ``location" required setting the edge (1, 3) type to ``has-author" (from ``has-location").

\begin{figure}[h]
\centering
\includegraphics[scale=0.45]{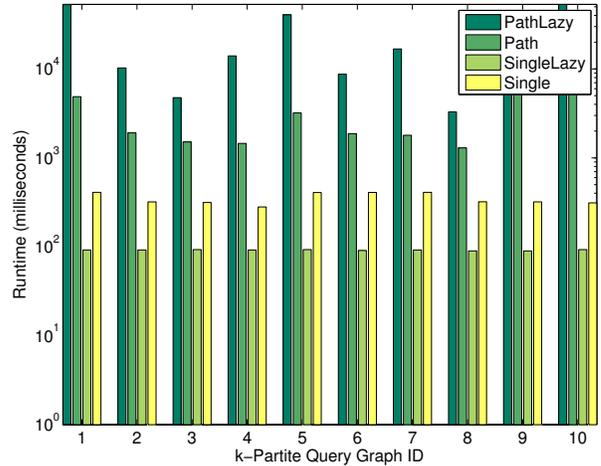}
\caption{Query processing times using four different search strategies for New York Times data.}
\label{fig:figs/query_planning/query_sweep_social_nyt_0-1M_0M_runtime_bar}
\end{figure}

\begin{figure}[h]
\centering
\includegraphics[scale=0.45]{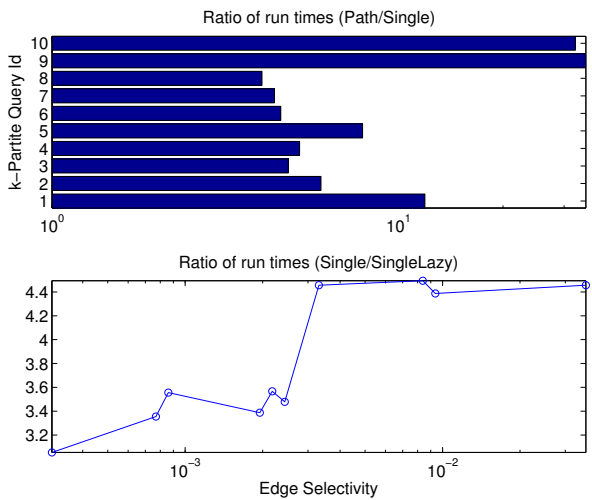}
\caption{Analyzing relative performance between the top-3 search strategies for New York Times data.  X-axis on the lower plot shows Expected Selectivity measured from edge distribution data.}
\label{fig:figs/query_planning/query_planning/nyt_speedup_details}
\end{figure}

Figure \ref{fig:figs/query_planning/query_sweep_social_nyt_0-1M_0M_runtime_bar} shows the total run time for processing 100,000 edges using each of these strategies.  Evidently, the "SingleLazy" strategy that combines lazy-search with 1-edge based decomposition is the best performing strategy.  We were surprised by the ``PathLazy" approach taking a disproportionate amount of time.  We discovered that for most of these queries, the ``PathLazy" approach resorted to searching for a ``twig" graph (example: an article connected to a person and a topic) and once it was found, it spawned off a search for another twig graph from all the matching vertices.  There are many high-degree vertices in the non-article (geo-location, organization, person, topic) category.  Therefore, any time the search finds a match around a high degree vertex such as ``geo-location:New York" or ``topic:Polictics and Government", it performs a second search.  The high runtime of ``PathLazy" results from a large number of occurrences of this event.  While these are insights drawn from deep analysis into the data, is there a generic way to determine which might be a better performing strategy?  We provide an answer in section \ref{subsec:Analysis via Relative Selectivity}.

Next, we investigate the relative performance of these strategies in more detail by studying the individual decompositions and degree distribution.  As we discussed in the time complexity analysis (section 4.1), the subgraph isomorphism cost for single edges is $O(1)$ and $O(\bar{d})$ for 2-edge subgraphs, where $\bar{d}$ is the average degree of the target vertex $v_2$ in the query subgraph.  To verify this, we plotted the ratio between the ``Path" and ``Single" strategies (Figure \ref{fig:figs/query_planning/query_planning/nyt_speedup_details}).  For the 2-edge or path based decomposition, 8 out of 10 queries were searching for a 2-edge subgraph that has an article vertex and other the two vertices chosen from the following combinations: topic and topic, organization and organization, person and person, geo-location and geo-location, topic and organization, topic and person, topic and geo-location, organization and person.  Two of the remaining decompositions resulted in searching for a 2-edge subgraph that has a \textit{person} connected to two distinct \textit{articles}, or a \textit{geo-location} connected to two distinct articles.  The average degree for vertices of type ``article" is 4.03.  As Figure \ref{fig:figs/query_planning/query_planning/nyt_speedup_details} shows, the speedup from ``Path" to ``Single" closely approximates the average degree of article vertices, as predicted in the complexity analysis.

Next, we observe the impact of adopting the Lazy approach for the ``Single" strategy.  The lower plot in Figure \ref{fig:figs/query_planning/query_planning/nyt_speedup_details} shows the speedup from Lazy Search as a function of \textit{Expected Selectivity} computed from edge distribution.  The speedup is seen to be higher with higher expected selectivity.  The expected selectivity is higher if the probability of individual edges appearing in the graph stream is high.  However, the probability of an edge's appearance in the graph stream does not provide us with any information about the presence of the query match.  Given the lazy approach guarantees the savings in search time, and consequently, reduces the number of partial matches being tracked, the savings are higher when the likelihood of the appearance of individual edges in the graph stream is high.

\begin{figure}[htbp]
\centering
\subfigure[]{\includegraphics[scale=0.35]{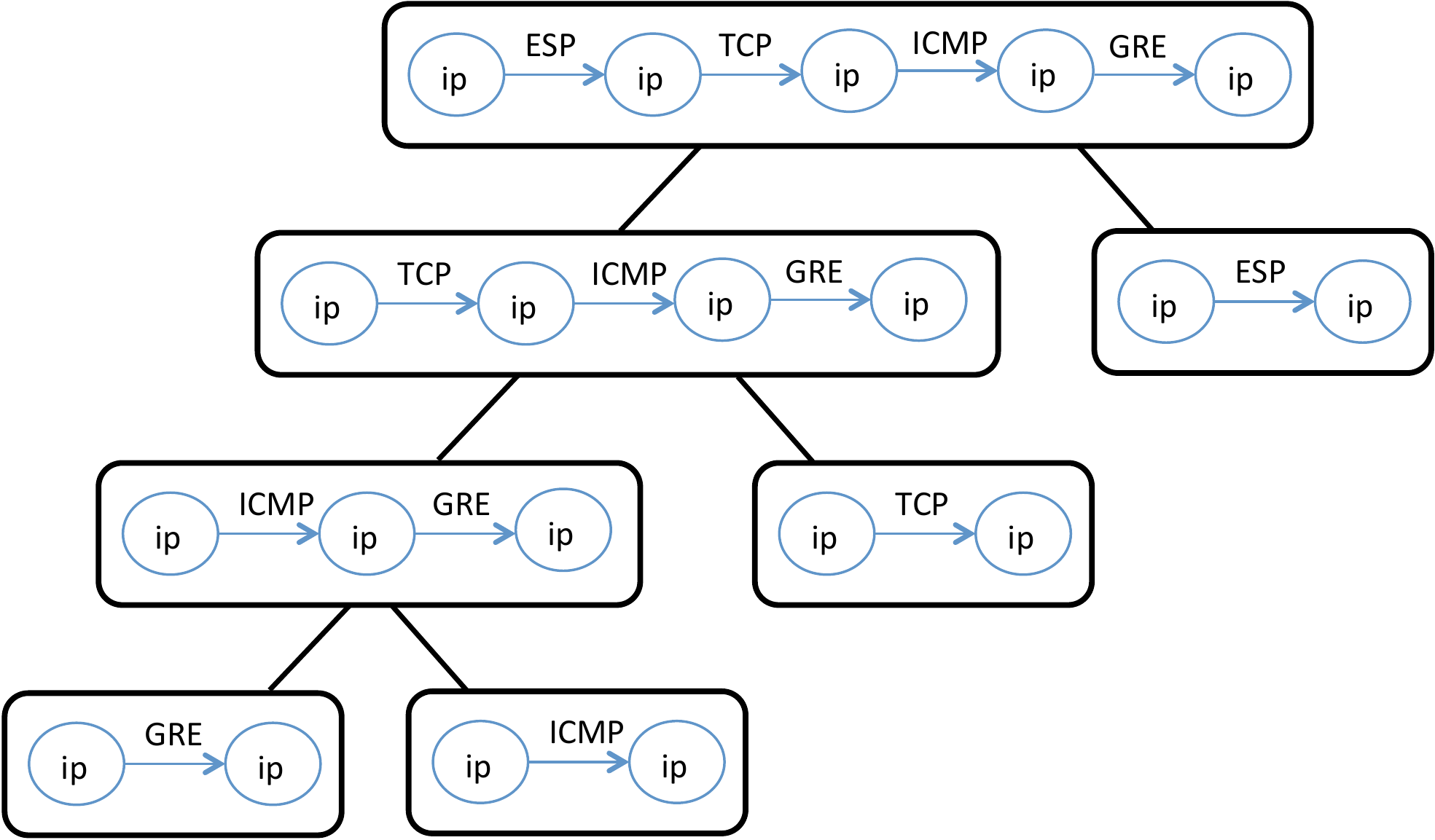}}
\subfigure[]{\includegraphics[scale=0.35]{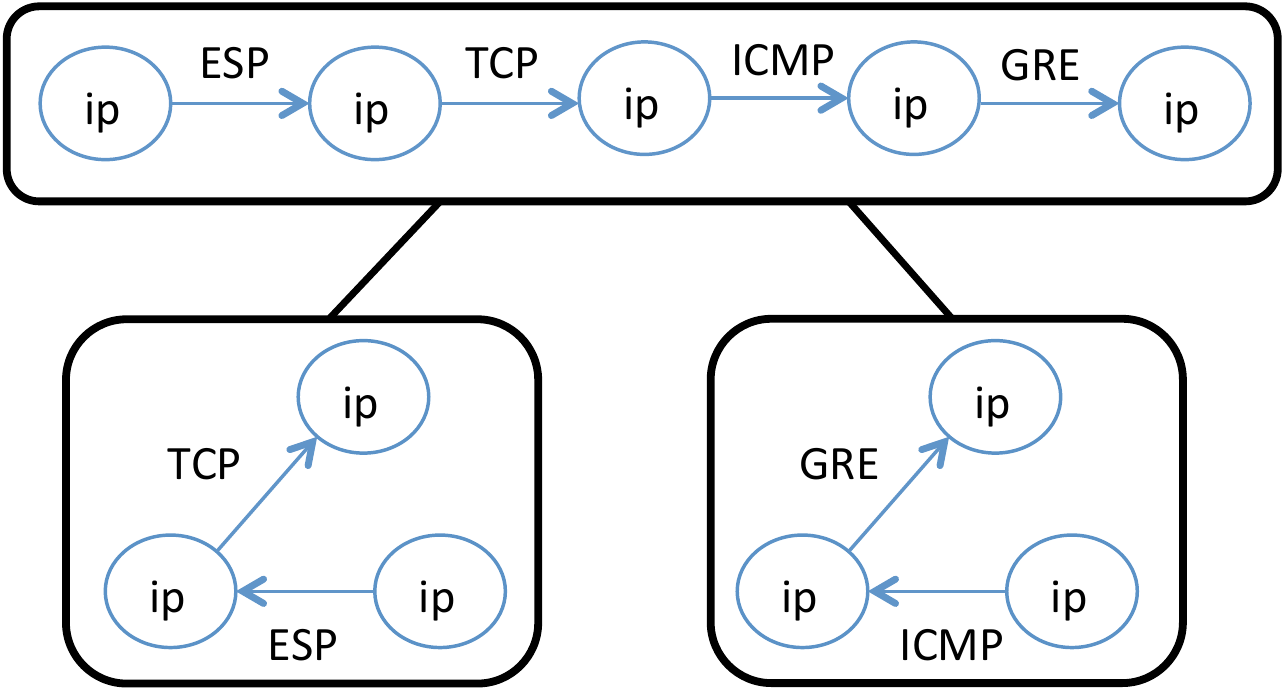}}
\caption{1 and 2-edge based decompositions of a path query on netflow traffic data.}
\label{fig:new_figs/apps/netflow_decomp}
\end{figure}

\begin{figure}[htbp]
\centering
\subfigure[Runtime for Path Queries on Netflow data.]{\includegraphics[scale=0.35]{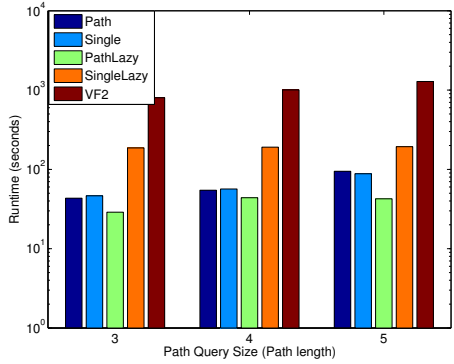}}
\subfigure[Runtime for Tree Queries on Netflow data.]{\includegraphics[scale=0.35]{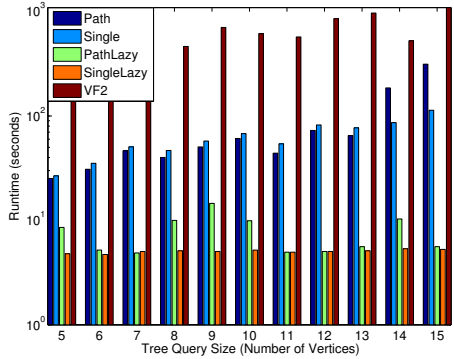}}
\subfigure[Runtime for Path Queries on LSBench data.]{\includegraphics[scale=0.35]{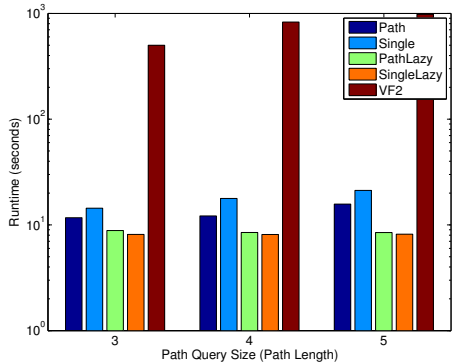}}
\subfigure[Runtime for Tree Queries on LSBench data.]{\includegraphics[scale=0.35]{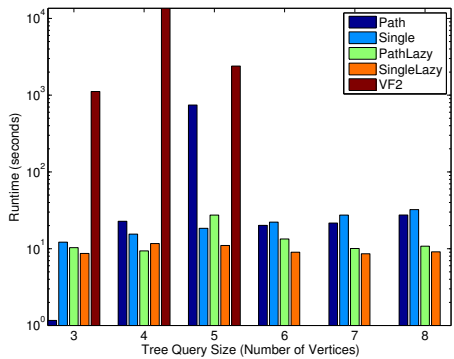}}
\caption{Runtimes from Path and Tree Queries on Netflow and LSBench.}
\label{fig:netflow}
\end{figure}

\begin{figure}[htbp]
\centering
\includegraphics[scale=0.45]{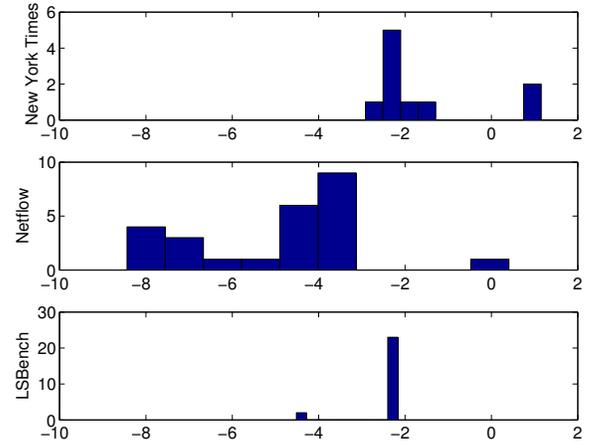}
\caption{Distribution of Relative Selectivity across queries with 4 edges in 3 datasets.  Relative selectivity is shown on X-axis in log scale.}
\label{figs/relative_selectivity_distribution}
\end{figure}

\subsubsection{Network Traffic and LSBench}

We present the results from the Netflow and LSBench dataset in this subsection.  Both of these datasets are orders of magnitude larger than New York Times and the scale allows us to magnify the differences between multiple strategies.

\textsc{Query Generation} We generate both path queries and binary tree queries for the netflow data.  Figure \ref{fig:new_figs/apps/netflow_decomp} shows two decompositions of an example query.  The  vertex labels are fixed to type ``ip" and the edge types are randomly chosen from a set of 7 protocols: ICMP, TCP, UDP, IPv6, AH, ESP and GRE.  The binary tree queries were generated following the test generation methodology described in \cite{DBLP:journals/pvldb/SunWWSL12}.   The LSBench dataset is tested with path queries and n-ary trees.  A list of valid triples (vertex type, edge type , vertex type)  is generated using the LSBench schema.  A tree query is generated by randomly selecting an edge from the set of valid triples and then iteratively adding valid new edges from any of the nodes available.  All our query graphs are unlabeled.  Using netflow data as an example, we do not generate a query that has a label associated with any of the nodes.  In practice, we expect users to employ labeled queries such as finding a tree pattern in the network traffic where the root of the tree has a IP address (i.e. label) from a certain subnet.  For social data, we may look for paths with specified user ids (node labels) on the source and the destination nodes on the path.  The impact of the selectivity of labels on query processing is explored
in our previous work  \cite{Choudhury:2013:DyNetMM}.  Here, our experiments are motivated to study the impact of subgraph distributional statistics on query processing.

\textsc{Comparison with Other Approaches}  In our previous work \cite{Choudhury:2013:DyNetMM} we had compared the performance of our implementation with the IncIsoMatch algorithm proposed by Fan et al. \cite{Fan:2011:IGP:1989323.1989420}.  Our IncIsoMatch implementation was based on a variant of the well-known VF2 algorithm \cite{cordella2004sub}.   Here, we compare our incremental algorithms with a non-incremental approach that performs subgraph isomorphism for the query graph (using VF2) on every new edge in the dynamic graph.

\textsc{Summarization of Results}  Unlike the New York Times dataset, where we reported the performance for each randomly generated query, we present aggregated results for each \textit{query group}.  All queries of the same type (path or tree) and size (3-hop length or 5 nodes) are denoted as a group.  We generated 100 queries for each group and then eliminated ones that contained 2-edge paths not seen in the sampled path distribution.  This was done for two reasons;  first, inclusion of an unseen 2-edge path combination makes the query artificially discriminative.  Our goal is to observe query processing time as a function of varying selectivity, so including unusually discriminative queries bias our studies.  Second, when asked to generate a path-based decomposition, our SJ-Tree generator resorts to generating a single-edge based decomposition when a query subgraph contains an unseen 2-edge path.  This would bias our comparison between a path-based decomposition and single-edge based decomposition.  Finally, for all the ``valid" queries we further sampled them by the Expected Selectivity computed using 2-edge path distribution and reduced each group to a smaller set of queries that provide a near uniform sampling of the Expected Selectivity from the larger set.  Finally, the reported runtime for a given strategy (e.g. ``PathLazy") is obtained by averaging the runtimes from the reduced set of queries,

Figure \ref{fig:netflow}a-d shows the query processing times collected for both datasets.  The size of the query processing window was fixed at 8M triples, and the performance statistics were collected at at the middle and at the end of the graph stream.  We profiled different components of the query processing such as the time spent in performing subgraph isomorphism and the time spent in updating the SJ-Tree.  The latter is largely composed of the time spent in looking up the hash tables in various nodes of the SJ-Tree, performing joins between partial matches and inserting new entries. We found that the subgraph isomorphism operation (for 1 or 2-edge subgraphs) dominates the processing time.  Considering both classes of queries with diameter 4 and 5, the subgraph isomorphism operation consumes more than 95\% of the total query processing time.

A general observation is that the performance of non-incremental search by VF2 is found to be 10-100x slower.  The Y-axis is plotted in log scale, and we can see how the run times of the ``Path" and ``Single" approaches rise exponentially as the query sizes are increased.  Overall, we find the ``SingleLazy" and ``PathLazy" are the best performing search approaches.   As the tree queries show, the growth rate in the query processing time is much slower for the ``Lazy" variants.  This conclusively demonstrates the effectiveness of restricting the search to where a match is emerging, and growing the match by starting from the most selective sub-query.



\subsection{Analysis via Relative Selectivity}
\label{subsec:Analysis via Relative Selectivity}

Figure \ref{figs/relative_selectivity_distribution} shows the distribution of relative selectivity for queries with 4 edges across all three datasets.  We picked query graphs with 4 edges to find a common basis for comparing different type of queries (k-partite vs. path queries) across multiple datasets, and the discussion is equally applicable to larger or different query class combinations.  The top subplot shows the relative selectivity of 10 k-partite queries from the New York Times data.  For netflow and LSBench, we randomly sampled 25 queries from the randomly generated path query collection.  As can be seen, the relative selectivity is very low for the netflow dataset.  Following the definition of relative selectivity, its value is lowered when the path distribution based selectivity is low.  In other words, there are some paths in the query which have very low probability of occurrence.  Therefore, the ``PathLazy" approach is superior for such queries.  Empirical observation on larger path queries and other tree queries seem to suggest two prominent clusters of relative selectivity values.  The first one typically ranges from 0.001 and above, and the second one contains values that are smaller by multiple orders of magnitude.  This suggests a heuristic that ``PathLazy" strategy could be employed for queries with relative selectivity below 0.001, and ``SingleLazy" be employed for queries above 0.001.
\section{Conclusion and Future Work}
\label{sec:Conclusion and Future Work}

We present a new subgraph isomorphism algorithm for dynamic graph search.  We developed a new data structure named Subgraph Join Tree (SJ-Tree) that represents the execution strategy for a query.  We also developed a set of algorithms for efficient collection of graph stream statistics, and using the statistics for automatic generation of the SJ-Tree for any given query graph.  We also introduced two selectivity metrics to quantitatively measure the hardness of a query by considering the temporal properties of the graph.  Given a stable distribution of edge types and the skew in the distribution of 2-edge subgraphs, we demonstrated that a query decomposition based on more selective edges will be consistently efficient.  We went further to introduce a ``Lazy" variant of the dynamic graph search algorithm that exploits the varying selectivity between different parts of a query graph.  We also extensively compared different search strategies, performing experiments on three different datasets and multiple query classes.  Finally, we concluded with a Relative Selectivity based rule for selecting a search strategy.


The paper investigates the important problem of estimating selectivity of query graphs.  While our 2-edge subgraph based approach provides an initial foundation, deeper investigations are warranted for more accurate selectivity estimation.  Subsequent research can leverage on the significant body of work on counting larger subgraphs such as triangles in streaming or semi-streaming scenarios.  Exploring query graphs with more diverse structures and developing a predictive model for accurate estimation of performance needs to be addressed.  Adaptive query processing is an important follow-up problem as well.  A long standing database query needs to be robust against shift in the data characteristics.  While we propose a fast algorithm for periodic recomputation of the primitive distribution, we do not address the issues of modeling the inefficiency from operating under a different selectivity order and migrating existing partial matches from one SJ-Tree to another.

\balance

\section*{Acknowledgment}
Presented research is based on work funded under the Center for Adaptive Supercomputing Software - Multithreaded Architecture at the US Department of Energy's Pacific Northwest National Laboratory, which is operated by Battelle Memorial Institute.
\bibliographystyle{abbrv}
\bibliography{citations_sutanay}
\balancecolumns
\end{document}